\input harvmac
\openup 1 \jot
\def\Title#1#2#3{#3\hfill \break \vskip -0.35in
\rightline{#1}\ifx\answ\bigans\nopagenumbers\pageno0\vskip.2in
\else\pageno1\vskip.2in\fi \centerline{\titlefont #2}\vskip .1in}


\def\R{\hbox{\rm I \kern-5pt R}}

\font\ticp=cmcsc10
\def\ajou#1&#2(#3){\ \sl#1\bf#2\rm(19#3)}

\lref\landaulifshitz{L.~Landau and E.~Lifshitz, {\it Quantum Mechanics},
Pergamon Press, Oxford (1977).}

\lref\grifflogic{R.B.~Griffiths, ``The Consistency of Consistent
Histories: A Reply to d'Espagnat'', Found. Phys., 23, 1601-1610, (1993).}

\lref\omnesa{R.~Omn\`es, ``About the Notion of Truth in Quantum Mechanics'',
J. Stat. Phys. 62, 841-861, (1991).}

\lref\omnesb{R.~Omn\`es, ``Consistent Interpretations of Quantum Mechanics'',
Rev. Mod. Phys. 64, 339-382, (1992).}

\lref\omnesc{R.~Omn\`es, {\it The Interpretation of Quantum
Mechanics}, Princeton University Press, Princeton, NJ (1994).}

\lref\gmhtwo{M.~Gell-Mann and J.B.~Hartle,
``Quantum Mechanics in the Light of Quantum Cosmology'', 
in {\it Proceedings of
the 25th International Conference on High Energy Physics, Singapore, August
2-8, 1990},
ed.~by K.K.~Phua and Y.~Yamaguchi (South East Asia Theoretical Physics
Association
and Physical Society of Japan) distributed by World Scientific, Singapore
(1990).}

\lref\gmhprd{M.~Gell-Mann and J.B.~Hartle, ``Classical Equations for Quantum
Systems'', Phys. Rev. D 47, 3345-3382, (1993).} 

\lref\gmhequiv{M.~Gell-Mann and J.B.~Hartle, ``Equivalent Sets of
Histories 
and
Multiple Quasiclassical Domains'', University of California Santa
Barbara preprint UCSBTH-94-09, gr-qc/9404013, submitted to Phys. Rev. D
(1994).} 

\lref\dowkerkentone{F.~Dowker and A.~Kent, 
``On the Consistent Histories Approach to Quantum Mechanics'',
J. Stat. Phys 82, 1575-1646, (1996).}

\lref\dowkerkenttwo{F.~Dowker and A.~Kent, ``Properties of Consistent 
Histories'', Phys. Rev. Lett. 75, 3038-3041, (1995).}

\Title{\vbox{\baselineskip12pt\hbox{ DAMTP/95-64}\hbox{quant-ph/9511032}{}
}}
{\vbox{\centerline {Remarks on Consistent Histories and 
Bohmian Mechanics${}^\dagger$}}}{~}
\vskip.2in
\centerline{{\ticp Adrian Kent}}
\vskip.1in
\centerline{\sl Department of Applied Mathematics and
Theoretical Physics,}
\centerline{\sl University of Cambridge,}
\centerline{\sl Silver Street, Cambridge CB3 9EW, U.K.}

\bigskip

\centerline{\bf Abstract}
{Recent work with Dowker on the scientific status of the 
consistent histories approach to quantum theory is reviewed and summarised. 
The approach is compared with formulations of quantum theory, such as 
Bohmian mechanics and the Copenhagen interpretation \`a la Landau-Lifshitz,
in which classical variables are explicitly appended.  
I try to explain why the consistent histories formulation
is scientifically problematic, in that it is a very weak theory, 
but also scientifically interesting, shedding new light on
quantum theory.}

\vskip2in

${}^\dagger~~$ Published in {\it Bohmian Mechanics and Quantum Theory:
An Appraisal}, J.~Cushing, A.~Fine and S.~Goldstein (eds), Kluwer
Academic Press (Dordrecht, 1996).  The article refers to 
discussions at the 1995 Bielefeld meeting, ``Quantum Theory 
Without Observers''.
\vfill\break

\footline={\ifnum\pageno=0 {\hfil} \else\hss\tenrm\folio\hss \fi}

A distressing feature of discussions of the problems of quantum theory
is their tendency to transform physicists from thoughtful and
sophisticated scientific critics into uncomplicated partisans or
unsympathetic spectators.  This holds true although the most
interesting questions posed by the various formulations of quantum
theory, and by rival theories, are precisely the type of scientific
and technical problems which physicists are trained to address.  What,
for example, does any given theory or formulation allow us to predict
or infer, and from what data?  Which of these predictions and
inferences can be tested?  How precise is the mathematical
formulation, and what mathematical properties does it have?  To what
extent is it consistent with important physical principles such as
invariance under Lorentz or general coordinate transformations?  How
elegantly is the theory formulated?  On how many arbitrary quantities
does it depend?

It would be good to reach consensus.  Our present lack of success
seems to stem less from subtle difficulties or metaphysical
differences than from the fact that scientific assessments of
interpretations of quantum theory or its rivals are unfashionable.  I
do not want to overstate the importance of scientific appraisal.  Of
course, creative theoretical work is the life blood of physics;
physicists need not, and perhaps generally should not, also be
philosophers of science.  But it is, surely, good to have a clear
understanding of what current approaches to quantum theory can, or
could possibly, achieve.  And, in fact, I would suggest that it is now
relatively easy to see that every approach to quantum theory leads to
serious problems and that there remain relatively few research
programs with any serious ambition of solving our difficulties.  This,
certainly, was the majority view at Bielefeld, though agreement on
precisely which are the serious programs and problems was harder to
come by.  A minority view, forcefully put at the meeting, was that the
problems of quantum theory have been solved by the consistent
histories formalism --- or at least that the form of a solution has
been convincingly sketched.  My impression at the end of the meeting
was that, although most of the participants believed this to be false,
those unpersuaded by consistent histories had reached no agreement on
precisely what constitute the problems --- still less the virtues ---
of the consistent histories program.

This article aims, no doubt optimistically, to explain both the
problems and the virtues, to persuade consistent historians that their
formalism is scientifically problematic, and to persuade sceptics that
it is nonetheless scientifically interesting.  I will try to show that
the consistent histories formulation, sensibly interpreted,
significantly changes our understanding of the scientific status of
quantum theory, not only because it offers a new formulation but also
because it sheds new light on earlier interpretations.  In particular,
I will compare and contrast the consistent histories approach, the
Copenhagen interpretation \`a la Landau-Lifshitz, and 
Bohmian mechanics.  
In so doing I will argue that some important technical 
claims made in the consistent histories literature turn out to be simply
false.  When sensibly interpreted, the formalism's 
chief virtue 
turns out to be not, as advertised, that it solves the 
problems of quantum theory, but rather that it highlights particular
scientific problems.  
Nonetheless, I will conclude, it can be used to give an interpretation of 
quantum theory which in 
important ways is 
better crafted, or at least more honest
about its deficiencies, than any of the standard interpretations.

This discussion of consistent histories is drawn from 
recent joint work with Fay Dowker.\refs{\dowkerkentone, \dowkerkenttwo} 
Our conclusions are set out at length in Ref. \refs{\dowkerkentone}.
Rather than repeating the details of the arguments here,
I will try to state the main results succinctly, to  
add some explanatory comments, and to respond to some
points raised during the Bielefeld meeting. 

Our thesis is the following.
We agree with other critics of consistent histories that 
the present interpretations of the formalism 
have some extremely unattractive features: indeed,
we show that in the cases of Omn\`{e}s and Gell Mann and Hartle
they have more serious problems.  
However, we distinguish between criticism of
the interpretations
offered in the literature and criticisms of
the consistent histories approach {\it per se}.
To take one example, 
Griffiths' proposal to interpret the formalism as defining a
non-classical logic is, of course, open to the usual
criticisms of quantum logic.  
This seemed to cause confusion at Bielefeld: some took Griffiths'
logic to be an essential part of the consistent histories program. 
Yet nothing in the consistent histories
formalism requires it.  
On the contrary, the formalism defines a sensible 
interpretation of quantum theory, using
ordinary logic and language, which we call the Unknown Set interpretation.   
It is instructive to examine this 
interpretation, to see
why it cannot be improved upon without going beyond the consistent 
histories formalism, and to understand its weakness as a scientific
theory --- for one is then
forced to appreciate both that the consistent histories formalism 
has virtues which other approaches to quantum theory lack, and that
it has defects which other formulations remedy. 

Let me now try to explain the reasons for our conclusions, 
assuming familiarity with the basic notions of consistent 
histories. 
For definiteness, consider the non-relativistic formulation in which sets of 
consistent histories are defined by sequences of projective 
decompositions $\{ \sigma_1 , \ldots , \sigma_n \}$ at times 
$t_1 , \ldots , t_n $, each $\sigma_i$ comprising projections 
$P_i^{1} , \ldots , P_i^{n_i}$, together with the Gell Mann--Hartle 
consistency conditions
\eqn\one{
Tr ( P_n^{a_n} \ldots  P_1^{a_1} \rho_i P_1^{b_1} \ldots P_n^{b_n} ) = 
\delta_{a_1 b_1} \ldots \delta_{a_n b_n} p(a_1 , \ldots , a_n ) \, , }
this last expression defining the probability of the history. 
Here $\rho_i$ is the density matrix defining initial conditions for 
the system, which I take to be the universe.  
There are other interesting formulations and consistency conditions in
the literature.  Moreover, the formalism admits a time-symmetric 
generalisation of quantum mechanics in which a second density
matrix $\rho_f$ defines final conditions. 
However, so far as we can tell, the basic scientific problems of 
the formalism are unsolved by any of these variations.

When the possible sequences of projective decompositions are suitably
parametrised, the Gell Mann--Hartle consistency conditions reduce to
simple algebraic equations in the parameters.  In other words, once
the boundary conditions are fixed, the classification
of consistent sets is a purely algebraic problem.  
It is hard to solve the relevant equations in any but the 
simplest of examples, or to prove general results about their solutions. 
Assume for the moment, though, that the equations have no very special
properties.  One would then expect that almost all the solutions
can be parametrised by a number of parameters equal to the number
of unknowns minus the number of consistency equations.  If the number
of parameters is much larger than the number of equations, one would
also expect that, given an approximate solution, one can generically
find an exact solution 
very close by.

I mention these mathematical trivialities because, if they apply to
consistent historical descriptions of real world physical 
events or experiments, they have interesting consequences for our 
understanding of the 
theory.\foot{
Unfortunately, though it would be surprising if this discussion did
not apply to real world physics, there seems to be no way to 
test the question directly.
It would, though, be interesting to 
test whether the consistency equations do indeed have the 
expected algebraic properties in moderately sized Hilbert spaces
and our conclusions hold in toy models.} 
Suppose, for example, we set up a series of $N$ independent experiments
in which (to use the standard Copenhagen language) 
distinct macroscopic devices measure observables of a 
microscopic quantum system and display the results by pointers, 
the experiments being complete and their results displayed 
at times $t_1 , \ldots , t_N$ separated by macroscopic intervals. 
Consistent historians can, of course, reproduce the standard 
probabilistic predictions for the results of these experiments, 
and do so roughly thus.  First, we identify the initial density matrix.
Next, we fix orthogonal projection operators at times $t_1, \ldots , t_N$,
corresponding to the possible positions of the relevant pointers and 
their complement. 
Then we argue that the decoherence effect of the environment (photons 
interacting with the pointers, and so on) will ensure that the 
set defined by these projections satisfies the consistency conditions 
to an extremely good approximation --- the off-diagonal terms, let us 
say, are no larger than $10^{-40}$.  Finally, we take this degree of
inconsistency as completely negligible, and simply use the standard
decoherence functional expressions for the probabilities of the various
results.  The justification for this procedure, given by 
Gell Mann and Hartle, is that we need not require a fundamental theory
to give precisely defined probabilities, or to give probabilities which
precisely obey the standard sum rules, since the 
purpose of theory is to calculate 
testable quantities and errors of $10^{-40}$ in 
our probability calculations are inconsequential
in any conceivable experimental test.  

If our expectations about the consistency equations are 
justified, we can improve on this discussion.  
The number of parameters needed to parametrise possible projection 
operators is hugely --- perhaps infinitely --- greater than the number of 
consistency equations here, and this has two important consequences.

The first is that the approximately consistent set used in 
these calculations could, in principle, be replaced by a very similar
exactly consistent set which would produce essentially the same 
probabilities --- indeed, we expect a parametrised family of
exactly consistent sets passing close to the set we initially used.  
We do {\it not} expect the sets in this family 
generally to involve projections we would naturally consider: their 
projections will generally be onto complicated subspaces of the 
Hilbert space describing the apparatus and its environment.  
Nor do we expect any single set in the family to be picked out in
any natural way.  
Nonetheless, the consistent histories formalism tells us that the 
family contains valid sets of histories with well-defined probabilities. 
Thus there is no need in principle ever to introduce approximately 
consistent sets: we can assume, without any serious fear of experimental 
contradiction, that exactly consistent sets are the 
only ones of fundamental physical relevance.  
Of course, this makes no practical difference, since we do not know 
precisely {\it which} exactly consistent set is relevant to any given 
experimental or cosmological calculation, and in practice --- unless and
until some rule is found which identifies the relevant consistent set 
for us --- we would generally use the usual approximately consistent set 
and accept that 
we thereby introduce small errors.  
Nonetheless, on this view, the formalism defines a mathematically precise
theory, and this --- if elegance
and lack of ad hockery are thought to be 
of any intrinsic merit --- must surely be counted a gain.  

The second and perhaps more significant consequence is that the consistent 
histories formalism shows that the standard Copenhagen description 
is chosen from a far larger class of possibilities than we previously
appreciated.  
For we expect the parametrised family of consistent sets
to be characterised by a very large number of parameters, and 
to be dominated by exactly consistent sets very far away from
the approximately consistent set we chose initially.  
Now it has certainly always been understood that there are slight
ambiguities in any Copenhagen description of a  
series of experiments, since there is a certain freedom in the choice 
of the Heisenberg cuts between system and observer.  
What is new here is the discovery that there is a continuous family 
of equally valid physical descriptions of the experiments and that,
while this family includes standard Copenhagen descriptions, 
almost all of its members involve variables quite different from,
and not even approximately deterministically related to, the
classical degrees of freedom used in the standard discussions.  
In the Copenhagen approach, any assignment of probabilities to
physical events not describable by classical degrees of freedom
is forbidden.  If we accept the consistent histories formalism
as a correct generalisation of the Copenhagen interpretation, we 
have to accept that such an assignment is theoretically sensible, 
and we then have to understand why the Copenhagen interpretation
is nonetheless all that we need for practical purposes.  

This, in fact, is the key question.  
The formalism offers a myriad of possible variables for 
describing physics.  Can it, suitably interpreted, explain
why the world reliably continues to appear to us always 
to be described by the particular measure zero subset corresponding
to familiar quasi-classical variables? 
Dowker and I argue that it cannot. 

In fact, we make the following stronger claim.  The scientific content
of the consistent histories formalism is given by the so-called 
Unknown Set Interpretation, which postulates that the fundamental
probabilistic theory of nature is defined by a choice of initial
density matrix, hamiltonian and canonical variables --- all of which
we might hope to specify precisely by some elegant theory --- together 
with some unknown and theoretically unspecifiable
consistent set of histories.  The histories from that set define the
sample space of possible events, and the decoherence functional then
defines the probability measure on that sample space in the usual way.
Thus, one history from the Unknown Set is chosen randomly to be 
realised, and it is this history which describes all of 
physics.\foot{It is often suggested that a fundamental theory
which assigns probabilities to a single event in this way is 
problematic, or even meaningless.  This is usually intended to 
be a criticism only of a particular type of theory, but seems 
in fact to imply 
a rejection of all probabilistic physical theories.  
For, practically speaking, a theory phrased in this way is no
more or less testable than any other probabilistic theory, since we
can only perform finitely many experiments.}
No other history, from this or any other set, is realised, and no set
other than the single Unknown Set is of any relevance for calculating
the probabilities of physical events.  

Let me emphasise at once that we are not suggesting that this
interpretation is ultimately satisfactory.  
We put it forward
to strip away what we see as inessential and sometimes confusing
proposals in the literature.  Our claim is that,
insofar as the consistent histories literature supplies sensible
interpretations of the formalism, those interpretations are almost
precisely 
scientifically equivalent to the Unknown Set Interpretation --- in
other words, they make almost precisely the same predictions, retrodictions
and inferences.  Where the literature claims to go beyond these
predictions, retrodictions and inferences, it is either erroneous or
else relies on significant assumptions extraneous to the consistent 
histories formalism.

Before making the case that the
Unknown Set interpretation really does encapsulate the scientific 
content of the formalism, let me discuss its scientific implications. 
An interesting question, raised at Bielefeld by Michael Dickson, is
whether so apparently weak an interpretation really deserves the title 
of a scientific theory.  To give a fully satisfactory answer would require
a general set of criteria for scientific theories.  
I do not have such a set of criteria.  However, it seems
to me that, once the initial density matrix, hamiltonian, and 
canonical variables are specified, the interpretation ought 
comfortably to pass any reasonable test.  
It is well-defined, and moreover comes from a quite elegant and
natural mathematical formalism; it is certainly falsifiable; one even, 
at present, has to admit (as a matter of logic rather than of plausibility)
the possibility that it is the best purely mathematical 
theory of nature which can be constructed.  
The interpretation makes one definite prediction, which is that
all the events we have observed to date, or will observe in the future,
can be described by a history of non-zero probability
from some consistent set.  If this fails to hold, as in principle it
could, then the interpretation --- and, of course, the entire 
consistent histories formalism --- must be rejected. 
The interpretation makes further predictions.  
These predictions are generally probabilistic, in the same way as 
the Copenhagen interpretation is. 
What is new and peculiar to the consistent histories formalism
is that they are also conditional on an unknown physical
quantity: the relevant consistent set. 
Given a particular observed history, we can predict that if a certain
event is described by one branch of a consistent extension of that
history, and if that particular consistent extension turns out to 
be part of the Unknown Set, then the relevant event will occur with
the conditional probability defined in the usual way by the 
decoherence functional.  If --- when, say, we attempt to predict 
the outcome of a series of experiments --- we find that we do
observe definite results (i.e., in this interpretation, that the 
corresponding projections do belong to the Unknown Set) but 
that the calculated conditional probabilities predict outcomes
significantly different from those observed then, again, we 
must reject either the theory being interpreted 
--- i.e. the specification of 
boundary conditions, hamiltonian, and canonical variables ---
or the interpretation itself.  
Likewise, even before we perform experiments, we are likely to 
reject some aspect of the theory if the observed history to date 
is highly improbable.  
Like most probabilistic tests, these last two of course require some  
intuitive or theoretical method of coarse-graining events. 
That is, we reject the theory not because a small probability event occurs
(all possible alternatives may have small probability) but because
we believe we can identify a natural division of the alternatives
into two classes and we find that sum of the probabilities of the events
in the class to which the occurring event 
belongs is small.  

The problem with the Unknown Set interpretation --- and, of course, the 
reason for doubt as to whether it constitutes a theory --- is that 
it gives no algorithm for making probabilistic predictions which depend only
on the observed data.  Every prediction takes the 
following form: ``{\it if~} the Unknown
Set contains the following projective decomposition at the following
future time, {\it then} the
probability for the future event 
described by one of the projections is $p$''. 
The interpretation does not predict that any future events will occur,
or that those which do occur will be describable in terms of 
familiar variables.  In particular, it does not predict that 
those variables which Gell-Mann and Hartle call quasi-classical ---
variables which describe macroscopic aggregates and which
generally follow deterministic equations of motion to a very good 
approximation --- will continue to be relevant.  Quite the contrary:
according to this interpretation, the apparent persistence of 
quasiclassicality is a great and inexplicable mystery.   
Thus, granted that the interpretation defines a scientific theory,
it is a theory with a glaring weakness.  
For most physicists, surely, believe that they {\it will} continue
to experience a quasiclassical world for the foreseeable future:
few are startled each morning by the dawning of yet another 
quasiclassical day. 
The persistence of quasiclassical experience, in other words, is part of  
our theory of nature.  Some well-known presentations of quantum
theory assume it explicitly.\refs{\landaulifshitz} 
Many do not, apparently because it has not been 
understood that there are well-defined interpretations 
of quantum theory in which quasiclassicality would not be perceived 
to persist, and that we need some scientific reason for rejecting
such interpretations.  A virtue of the consistent histories 
formalism, in the Unknown Set interpretation,
is that it makes these points absolutely clear.   

Let us now turn to the interpretations in the consistent histories
literature.  It is only possible to outline the arguments here, but 
perhaps a brief precis will be of use.  I hope the reader, and those 
criticised, will forgive the necessarily crude summaries.  

Griffiths\refs{\grifflogic} suggests 
that the consistent histories formalism
should be interpreted as defining a new logic adapted to propositions
describing the physical world.  Griffiths' logic has the property that
any two propositions referring to projections belonging to different
consistent sets can be true without implying that their conjunction
is true.  We can, for instance, predict 
that the detectors at CERN will function tomorrow in the ordinary way,
producing quasiclassical records of the events they detect, and also
predict that the detectors, and their recording devices, and much else
besides, will {\it not} behave quasiclassically tomorrow.  We cannot, 
however, use Griffiths' logical rules to
deduce the prediction that the detectors both will and will
not behave quasiclassically.  
We hence avoid contradiction 
though --- as is usual with quantum logic ---
at the price of a theory which we simply do not understand how to
interpret.  Griffiths' interpretation, however, skirts the key
point.  We can never experience the truth or falsity of 
propositions from more than one consistent set.  If the formalism
is fundamentally correct then all our scientific
endeavours will be described by one consistent set 
and the scientifically relevant problem is the identification of that set. 
We can, of course, do calculations in other sets; we 
can too, if we wish,
manipulate propositions involving other sets according to Griffiths'
logical rules --- but neither of these activities are of any use
in predicting the future we will actually experience. 
Griffiths' interpretation is scientifically equivalent to
the Unknown Set interpretation. 

Omn\`{e}s also interprets the formalism as defining rules for the
logical analysis of propositions about the physical 
world.\refs{\omnesa, \omnesb, \omnesc}
Omn\`{e}s' logics are conventional: propositions belonging
to incompatible consistent sets simply cannot be discussed together.
The significant new proposal in Omn\`{e}s' interpretation is the
notion of a ``true proposition'' --- a proposition which is not
given to us in the form of observed data, but is deducible from
those data by a new rule appended to the consistent histories formalism.
Unfortunately,\refs{\dowkerkentone} as Omn\`{e}s accepts, the rule he
originally proposed fails to allow the intended deductions: indeed,
it generally seems to allow almost no 
deductions.\foot{Could another definition of ``truth'' 
do the job?  Omn\`{e}s has new proposals.  Dowker and I too have 
investigated possible alternative rules.  
Our tentative conclusion is that interesting rules do exist
which allow at least some non-trivial inferences about the past, 
but we can identify no rule which allows any useful predictions of the
future.  We hope to give a detailed discussion elsewhere.}  

Perhaps it is a slight overstatement to say that we are 
left with an interpretation scientifically equivalent to the
Unknown Set interpretation: this depends on exactly how 
narrowly one defines science when its subject matter is the past. 
Any principle which allows even a few inferences about the 
past, untestable though they may be, would probably generally be regarded
as scientifically useful if those inferences form part of an elegant
and compelling theoretical explanation of present data.  
There seems to be no evidence that any inferences implied by Omn\`{e}s' 
original criterion do so, but the possibility cannot be completely excluded. 
However, so far as the criteria for truth in the existing literature are 
concerned and insofar as they apply to predictions, 
we must indeed conclude that they indeed
do not affect the scientific status of the formalism.  

Gell Mann and Hartle's 
conclusions,\refs{\gmhtwo, \gmhprd, \gmhequiv} however, certainly 
go beyond those implied by the Unknown Set interpretation.  
In fact, Dowker and I argue\refs{\dowkerkentone} that 
those conclusions are not entirely
coherent in their use of the formalism.  
Nonetheless, the central claim of Gell Mann and Hartle's interpretation
is tenable.  
This is the suggestion that quasiclassicality appears to us to persist 
not because quasiclassical variables play any special role in 
the theory, but because we ourselves have evolved organs of perception
which are sensitive to those variables and a mental apparatus which
represents the world in quasiclassical terms.   
But is this a valid argument without further assumption, or
is it, like other recent ideas relating consciousness to quantum theory, 
a speculation?  

At first sight it appears not only a valid argument but 
close to a truism.  
Almost all scientists would agree that 
our perceptions and our 
mental algorithms have evolved to become highly sophisticated 
at gathering and utilising quasiclassical data.  
This agreement, though, is predicated on the assumption that
one may assume a quasiclassical description of the world.
Quasiclassical variables arise naturally in higher 
order theories of nature such as classical mechanics, chemistry, and 
terrestrial biology, and evolutionary biologists take their use
for granted. 
Likewise, our theories of brain function are 
classical theories and our understanding of consciousness, such as it is,
is entirely based on classical models. 
We cannot use biological science to justify any
general conclusions about evolution, perception,
or consciousness from within a novel interpretation 
of quantum theory such as the consistent 
histories interpretation.  
For if we take seriously a theory --- such as the consistent 
histories formalism --- which 
describes us as being in
superpositions of quasiclassical states, or in states defined 
in terms of entirely non-quasiclassical variables,
we can make no statement about our perceptions in those 
states without new hypotheses. 
Such hypotheses would necessarily be speculative: they certainly do
not follow from our conventional, quasiclassical
understanding of the relation of perception to brain function; nor
do they follow from any empirical data or theoretical 
insight presently available to us.  
What one would need, in fact, is a theory of consciousness written directly 
in the language of the formalism.  It is hard to imagine, and Gell-Mann and 
Hartle do not try to explain, how one would presently go about trying to 
formulate such a theory.  

We conclude, then, modulo a minor caveat about Omn\`{e}s' treatment of 
the past, that the Unknown Set interpretation 
is indeed scientifically equivalent to the interpretations 
discussed in the literature, when they are stripped of 
extraneous hypotheses.
Although there is currently no single canonical formulation of 
quantum theory, it can reasonably be argued 
that the consistent histories formulation
is the minimal formulation of the quantum theory of a closed system 
which produces a well-defined 
scientific theory.  It tells us that, even 
when we ignore general relativity, our 
theory of the macroscopic world --- and in particular our 
expectation of its persisting quasiclassicality --- involves assumptions
that go beyond both quantum theory and any theory of the cosmological
boundary conditions.  
Some thoughtful 
critics and advocates of orthodox quantum theory have long
appreciated this.  Perhaps it has remained 
controversial only through the wider confusion over 
interpretations of quantum theory.  
The consistent histories formalism now so clearly defines 
a natural interpretation of quantum theory, and spells the 
conclusion out so precisely, that it is hard
to see how any serious controversy can persist. 

As we have seen, the formalism also shows that there are equally valid,
perfectly well-defined, alternatives in which quasiclassicality persists
only for an interval, or never arises. 
This is an important 
new development in our understanding of quantum 
theory.\foot{Its novelty might be disputed.  
It is true that something similar occurs in interpretations of quantum 
mechanics in which the events at different times are entirely
uncorrelated.  For such interpretations arbitrary basis selection 
rules can be used at each point in time, and in particular one can 
use rules in which the system lies in an eigenstate of quasiclassical
operators for a while and an eigenstate of non-quasiclassical
operators thereafter.  But few take such ahistorical interpretations
seriously.  
One can probably find historical interpretations, 
other than the consistent histories formalism, in which quasiclassicality
does not persist --- for example, it ought to be possible to 
produce generalised Bohmian theories with this property ---
but I can think of no discussion of such interpretations in the 
literature.}  

Bohmians and collapse model theorists may happily accept that the formalism
supplies new arguments against orthodox quantum theory,
but will perhaps feel our analysis confirms their belief that the
formalism has no positive scientific use.  
Is the formalism not, after all, scientifically sterile? 
Were the problems it illustrates not solved long ago by Bohmian 
mechanics?  Are dynamical collapse models, with their intriguing alternative
explanation of quasiclassicality, not a far more vital subject of research?  
I sympathise with the spirit of these questions, but let me end by 
explaining why I cannot dismiss the formalism so conclusively.

First, while defining the problems of quantum theory very
clearly, it also suggests an interesting possible form of a solution.
All one needs is a rule (perhaps probabilistic) which takes as input
the dynamics and boundary conditions of a theory and produces as
output a consistent set (the Unknown Set) which turns out 
to be quasiclassical.  This, of course, begs the question of whether
such a rule can be found.  
Here the superiority of Bohmian mechanics
and of GRW-type collapse models is presently clear, since the analogous 
selection principle is already known in both cases and both theories
explain quasiclassicality.  It remains to be seen
whether they give the right explanation and whether they are 
capable of giving any explanation in the context of relativistic 
quantum field theory: the same, of course, is true of the 
consistent histories formalism.   

Second, the consistent histories formalism seems to be
a strong competitor theory where cosmological
applications are concerned.  For example, one can easily imagine theories
of structure formation involving a series of past events
which can be described within a consistent set but which, 
even if a good Bohmian cosmological theory were to exist, 
could not naturally be described in terms of Bohmian trajectories. 

Third, the consistent histories formalism surely 
ought to be explored further 
precisely because it is a good formulation of quantum theory.  
If the eventual goal is to go beyond quantum theory, it is probably as
well to understand all interesting formulations and 
interpretations of the theory.  
Different formulations, after all, may inspire
different post-quantum theories. 
We need, in particular, to 
understand the possible definitions of ``truth'' and 
their properties; a general treatment of  
quantum field theory in the consistent histories 
formulation; and a clearer understanding of how
the formalism applies to cosmology.  

\vskip1in

\leftline{\bf Acknowledgements}  
I am very grateful to Fay Dowker for the collaboration 
reported here and for many discussions of the
consistent histories approach.  This work was supported by a Royal
Society Research Fellowship. 

\listrefs
\end